\documentclass[%
 reprint,
 amsmath,amssymb,
 aps,
]{revtex4-2}

\usepackage{graphicx}
\usepackage{dcolumn}
\usepackage{bm}

\begin{document}
\title{Arbitrage Equilibrium, Invariance, and the Emergence of Spontaneous Order \\
in the Dynamics of Birds Flocking}

\author{Abhishek Sivaram}
\affiliation{
Complex Resilient Intelligent Systems Laboratory, Department of Chemical Engineering, Columbia University, New York, NY 10027, U.S.A.}

\author{Venkat Venkatasubramanian}%
\email{venkat@columbia.edu}
\affiliation{%
Complex Resilient Intelligent Systems Laboratory, Department of Chemical Engineering, Columbia University, New York, NY 10027, U.S.A.}

\date{\today}

\begin{abstract}
The physics of active biological matter, such as bacterial colonies and bird flocks, exhibiting interesting  self-organizing dynamical behavior has gained considerable importance in recent years. Recent theoretical advances use techniques from hydrodynamics, kinetic theory, and non-equilibrium statistical physics. However, for biological agents, these don't seem to recognize explicitly their critical feature, namely, the role of survival-driven \textit{purpose} and the attendant pursuit of \textit{maximum utility}.  Here, we propose a novel game-theoretic framework and show a  surprising result that the bird-like agents self-organize dynamically into flocks to approach a \textit{stable arbitrage equilibrium} of equal effective utilities. While it has been well-known for three centuries that there are constants of motion for passive matter, it comes as a surprise to discover that the dynamics of active matter populations could also have an invariant. What we demonstrate is for ideal systems, similar to the ideal gas or Ising model in thermodynamics.  The next steps would involve examining and learning how real swarms behave compared to their ideal versions. Our theory is not limited to just birds flocking but can be adapted for the self-organizing dynamics of other active matter systems.
\end{abstract}
\keywords{Active matter, Arbitrage equilibrium, Invariance, Self-organization, Statistical teleodynamics, Flocking, Emergence}
\maketitle



\section{Introduction}
Active matter describes systems composed of large numbers of self-actualizing agents, which consume and dissipate energy resulting in interesting dynamical behavior \cite{toner2005hydrodynamics, narayan2007long, ramaswamy2010mechanics,marchetti2013hydrodynamics, cates2013active, cates2015motility,  gonnella2015motility, berkowitz2020active, o2020lamellar}. Biological examples of such systems include self-organizing bio-polymers, bacteria, schools of fish, flocks of birds, and so on. This study is focused on birds flocking. 

Flocking has been studied extensively from dynamical systems and statistical mechanics perspectives~\cite{reynolds1987flocks, vicsek1995novel, ballerini2008interaction, bialek2012statistical, bialek2014social}. Such analyses have contributed substantially to our evolving understanding of interesting emergent properties such as phase segregation, flock stability, etc. However, these approaches don't seem to model \textit{explicitly} the critical feature of active biological agents, namely, the role of \textit{purpose} and its attendant  \textit{pursuit of maximum utility}. Being biological agents, birds are innately purposeful, driven by the goal to survive and thrive in challenging environments as Darwin explained. We believe any comprehensive theory of active biological matter has to overtly account for this defining characteristic of the agents. 

We address this need by using a novel game-theoretic framework, which we call \textit{statistical teleodynamics} \cite{venkat2015, venkatasubramanian2017statistical, venkat2017, venkatasubramanian2019much, kanbur2020occupational, venkatasubramanian2022unified}. The name comes from the Greek word \textit{telos}, which means goal. Just as the dynamical behavior of gas molecules is driven by thermal agitation (hence, \textit{thermo}dynamics), the dynamics of purposeful agents is driven by the pursuit of their goals and, hence, \textit{teleo}dynamics. Statistical teleodynamics may be considered as the natural generalization of statistical thermodynamics for \textit{purpose-driven} agents in active matter. It is a synthesis of the central concepts and techniques of potential games theory with those of statistical mechanics towards a unified theory of emergent equilibrium phenomena in active and passive matter \cite{venkatasubramanian2022unified}. 

In this paper, we study a model of birds flocking that is inspired by the well-known Reynolds's Boids model, analytically and computationally. In the next two sections, we briefly review the dynamical systems and statistical mechanics perspectives, respectively. This is followed by our statistical teleodynamics formulation, which starts with a quick introduction to potential games. We conclude with a discussion of the main results and their implications.  

\section{Dynamical Models of Flocking}

Flocking has been studied using dynamical models, which describe the time evolution of the position $\mathbf{r}_i$ of the $i^{\text{th}}$ agent and its velocity $\mathbf{v}_i$ using pre-specified rules for change. The state of the flock at any given time is specified by specifying $\mathbf{r}_i$ and $\mathbf{v}_i$ for all agents. When the agents move at a constant speed $v_0$, the state of the system is then determined by the set of agents' positions and velocity directions or orientations $\{\mathbf{r}_i, \mathbf{s}_i\}_{i=1}^N$, where $N$ is the number of agents.  Both the Reynolds's model \cite{reynolds1987flocks} and the Vicsek model \cite{vicsek1995novel} describe the time evolution of an agent's velocity, but using different force models. 

An agent $i$ is said to be affected an agent $j$, if $j$ is in the neighborhood of $i$, $\mathcal{N}^i$. The neighborhood $\mathcal{N}^i$ of $i$ is defined by a matrix whose elements are $n_{ij}$, where
\begin{eqnarray}
    n_{ij} = \begin{cases}
      1 & \text{if $j$ is a neighbor of $i$}\\
      0 & \text{otherwise}\\
    \end{cases}    
\end{eqnarray}

The span of the neighborhood is specified in terms of the absolute distance between $i$ and  $j$, and a size parameter $r_0$ \cite{vicsek1995novel, gregoire2004onset}, such that

\begin{eqnarray}
    n_{ij} = \begin{cases}
      1 & |\mathbf{r}_i^{(t)} - \mathbf{r}_j^{(t)}| < r_0\\
      0 & \text{otherwise}\\
    \end{cases}    \label{eq:neigh}
\end{eqnarray}

As seen from Eq. \ref{eq:neigh}, we consider agent $i$ to be its own neighbor. One can also define a neighborhood in terms of a fixed topology of nearest neighbors \cite{ballerini2008interaction, bialek2012statistical, bialek2014social}, but we don't use this specification in this study. It follows that the number of neighbors of an agent $i$ is given by $n_i = \sum_j n_{ij}$.

In the Reynolds's model, the agents (called \textit{boids}) obey the following three rules as they fly around:
\begin{enumerate}
    \item \textit{Rule of cohesion}: A boid steers to move towards the average position of local flockmates
    \item \textit{Rule of separation}: A boid steers to avoid collision and crowding of local flockmates
    \item \textit{Rule of alignment}: A boid steers towards the average heading of local flockmates
    
\end{enumerate}





In general, we can write the net effect of these forces on the $i^{\text{th}}$ boid by the equation (Eq. \ref{eq:boids_dt}), 
\begin{widetext}
\begin{eqnarray}
    \mathbf{v}_i(t+\Delta t) = \mathbf{v}_i(t) + \left(\frac{a}{n_i} \sum_{j} n_{ij} (\mathbf{r}_j - \mathbf{r}_i) + b \sum_{j} n_{ij} (\mathbf{r}_i - \mathbf{r}_j) \frac{c}{n_i} \sum_{j} n_{ij} (\mathbf{v}_j - \mathbf{v}_i) + \boldsymbol{\eta}(t)\right ) \Delta t
    \label{eq:boids_dt}
\end{eqnarray}
\end{widetext}

where $a, b,$ and $c$ are parameters corresponding to the rules of \textit{cohesion, separation}, and \textit{alignment}, respectively. Parameter $\boldsymbol{\eta}$ is the uncorrelated noise in the agent's velocity.
 The time-scale, $\Delta t$, in Eq. \ref{eq:boids_dt} can be subsumed in $a, b, c$. {The Vicsek model, similarly, updates velocity purely as a function of the alignment of the agent with its neighbors, though modifications have been proposed to include pairwise forces \cite{bialek2012statistical, gregoire2004onset} (see also, Supplementary Information, Section 2).}

It's important to note that in both the Reynolds model and the Viscek model, there is no concept of the ``final" state, or an equilibrium state, of the system as time tends to infinity. In this regard, they are like molecular dynamics simulations of molecules where there is no concept of a final  equilibrium state in the equations. They can only determine the immediate next move of the molecules, at any given time, not their final configurations. The final outcome is determined, \textit{a posteriori}, after the simulations are run for a long time. We highlight this important point here as our theory differs conceptually in this regard. 

{While dynamical models of flocking have been studied extensively in literature, work has also been reported using statistical mechanics to analyze the flocking behavior. These methods typically use a maximum entropy formulation via an Ising-model inspired Hamiltonian of the boids' interaction \cite{bialek2012statistical, bialek2014social}}.

\section{Statistical Teleodynamics of Flocking -- A Game-theoretic Formulation}

The Reynolds and Vicsek models specify bottom-up agent-level dynamical behavior, but they don't provide an analytical framework to predict the dynamics of the entire flock. This is determined only computationally via agent-based simulations. That is, there is no analytical framework to derive the behavior of the whole from the behaviors of the parts. 

On the other hand, the statistical mechanics formulation \cite{bialek2012statistical} is a top-down approach that starts with the specification of the Hamiltonian of the flock and then imposes the maximum entropy distribution on it. It is not clear why maximum entropy, which is obviously relevant for passive matter systems, would be applicable for survival-driven birds. The typical statistical mechanics approach uses the superficial similarity between spins in magnetic systems (e.g., the Ising model) and the orientation of the birds to apply maximum entropy methods. The deeper question of why this is conceptually relevant for the birds is not addressed. Most importantly, all these approaches don't seem to recognize \textit{explicitly} that active agents such as birds act \textit{instinctively} to improve their survival prospects.  

We address these challenges using our \textit{statistical teleodynamics} framework \cite{venkat2015, venkatasubramanian2017statistical, venkat2017, venkatasubramanian2019much, kanbur2020occupational, venkatasubramanian2022unified}. In this theory, the \textit{fundamental} quantity is an agent's \textit{effective utility} which is a measure of the net benefits pursued by the agent. Every agent behaves \textit{strategically} to increase its effective utility by switching states and exploiting \textit{arbitrage} opportunities.  

In our theory of flocking, we propose that birds are \textit{arbitrageurs} that always maneuver to increase their effective utilities, which determine their survival prospects, dynamically in flight. The effective utility of a bird depends on its position, speed, and alignment with the rest of the members in its neighborhood. 

Thus, we interpret the three rules of engagement for Reynolds's boids not as externally imposed forces on the agents but as \textit{innate}, self-actualizing, properties of the agents acquired over millions of years of Darwinian evolution. These instinctive characteristics enable the agents to incessantly search for better effective utilities in order to improve their survival chances. 

Hence, we believe that the proper formulation of flocking ought to start with a model of effective utility that a bird uses to make such decisions dynamically in flight. Seen from this perspective, we suggest that birds do not fly \textit{randomly} (as statistical mechanics-based formulations implicitly assume), but maneuver \textit{strategically} to improve their utilities. We exploit this critical insight to model the dynamical behavior of birds in flight by using the concepts and techniques from potential games. 

In potential games, there exists a single scalar-valued global function, called a {\em potential} ($\phi(\mathbf{x})$) that has the necessary information about the payoffs or the utilities of the agents. The {\em gradient} of the potential is the utility, $h_i$, of the $i$th agent ~\cite{rosenthal1973class,sandholm2010population, easley2010networks, monderer1996potential, venkat2015}. Therefore, we have 

\begin{equation}
{h}_i(\mathbf{x})\equiv {\partial \phi(\mathbf{x})}/{\partial x_i}
\end{equation}

where $x_i=N_i/N$ and $\mathbf{x}$ is the population vector. A potential game reaches equilibrium, called \textit{Nash equilibrium} (NE), when the potential $\phi(\mathbf{x})$ is maximized. Furthermore, this Nash equilibrium is unique if  $\phi(\mathbf{x})$ is strictly concave \cite{sandholm2010population}. At Nash equilibrium, all agents enjoy the same effective utility, i.e., $h_i = h^*$. In fact, the equality of effective utilities in all states is the \textit{fundamental criterion of game-theoretic equilibrium} for active matter. It is an \textit{arbitrage equilibrium} \cite{kanbur2020occupational} where the agents don't have any incentive to switch states anymore as all states provide the same effective utility $h^*$. Thus, the maximization of $\phi$ and $h_i = h^*$ are exactly equivalent criteria, and both specify the same outcome, namely, an arbitrage equilibrium. 

There is a deep and beautiful connection between potential game theory and statistical mechanics as discussed by Venkatasubramanian \cite{venkat2017, venkatasubramanian2022unified}. Since an 
elaborate discussion about this would take us afar from the objectives of this paper, we refer interested readers to \cite{venkat2017, venkatasubramanian2022unified} . 

\subsection{Garud's Utility: Position Dependence}

Our goal here is to develop a simple model of the effective utility ($h_i$) of our  boid-like agent, called \textit{garud} (after the legendary king of birds, \textit{Garuda}, in Indian mythology). 

We want the model to be an appropriate coarse-grained description of the system that can make useful predictions not restricted by system-specific nuances. We have tried, deliberately, to keep the model as simple as possible without losing key insights and relevance to empirical phenomena. One can add more complexity as and when desired later on. What we are aiming for is the equivalent of the ideal gas model or the Ising model for birds flocking. 

We develop our teleodynamical model using Reynolds's model as the start, but our approach is not restricted to this example alone; it is applicable to other models as well. 

We consider a garud's position in the frame of reference of the center of mass of its neighborhood. We then apply the \textit{rule of cohesion} and \textit{rule of separation} to formulate the model for utility. 

The rule of cohesion requires the garuds to come together and hence an $i^{\text{th}}$ garud's utility increases as it has more neighbors, $n_i$. However, the increased utility comes at the cost of \textit{congestion}, the \textit{disutility of congestion} (corresponding to the rule of separation). The trade-off between the two terms, the benefit-cost trade-off, results in an inverted-U profile, which, following Venkatasubramanian ~\cite{venkat2017}, can be parameterized as,
\begin{equation}
    h_r^{(i)} = \alpha n_i - \beta n_i^2 \label{eq:ur}
\end{equation}

where $h_r^{(i)}$ is the position component of the utility for the $i^\text{th}$ $\alpha, \beta >0$. Note that the positional dependence is accounted for in the computation of $n_i$. Given a configuration of $\{\mathbf{r}_i\}$, the neighborhood of the $i^{\text{th}}$ garud is defined by the parameter $r_0$, where if $j^{\text{th}}$ garud is within this radius then it is considered a neighbor.

This in turn identifies a direction of increased utility, given by,
\begin{equation}
    \frac{\partial h_r^{(i)}}{\partial \mathbf{r}_i} = \alpha \frac{\partial n_i}{\partial \mathbf{r}_i}  - 2\beta n_i \frac{\partial n_i}{\partial \mathbf{r}_i}  \label{eq:ur-der}
\end{equation}

$\dfrac{\partial n_i}{\partial \mathbf{r}_i}$ is dependent on the garuds in the perimeter of the neighborhood of the reference garud $i$.

\subsection{Garud's Utility: Velocity Dependence}

The utility of a garud is also dependent on the velocity of its neighbors in that the garud attempts to match the orientation with its neighboring garuds. This utility component ($h_v^{(i)}$) can be written as,
\begin{equation}
    h_v^{(i)} =   {\gamma} \sum_{j} n_{ij} \frac{\mathbf{v}_i}{|\mathbf{v}_i|} \cdot \frac{\mathbf{v}_j}{|\mathbf{v}_j|}
\end{equation}

The utility of the $i^{\text{th}}$ garud, then, depends on the orientation of the other garuds in its neighborhood, i.e., $\mathbf{s}_i\cdot \mathbf{s}_j$ where $j$ is a neighbor of garud $i$. This gives the \textit{alignment utility} for the $i^{\text{th}}$ garud as,
\begin{equation}
    h_v^{(i)} = \gamma \sum_{j} n_{ij} \mathbf{s}_i\cdot\mathbf{s}_j 
\end{equation}

where $n_{ij}$ shows if the $j^\text{th}$ garud is a neighbor of garud $i$, i.e., 
\begin{eqnarray}
    n_{ij} = \begin{cases}
      1 & |\mathbf{r}_i{(t)} - \mathbf{r}_j{(t)}| < r_0\\
      0 & \text{otherwise}\\
    \end{cases}    
\end{eqnarray}

If each garud is perfectly aligned with its neighbors, this utility component is maximal, whereas if they are oriented in the opposite direction this is minimal. Therefore, the garuds prefer to be aligned. This gives the $i^{\text{th}}$ garud an arbitrage opportunity to adjust its velocity vector towards this direction to increase its utility. This opportunity for increasing its utility generates a self-propelled force on the $i^{\text{th}}$ garud. If  the $i^{\text{th}}$ garud is not aligned with its neighbors, this direction of increased utility is given by,

\begin{equation}
    \frac{\partial h_v^{(i)}}{\partial \mathbf{s}_i} = \gamma \sum_{j} n_{ij}\mathbf{s}_j
\end{equation}

\subsection{Garud's Effective Utility}

There is one other utility component remaining to be considered. This is not stated explicitly in the three rules of the boids. However, it is implied because it is assumed that the boids have to be moving constantly.  

So, as a garud incessantly moves and jockeys for better positions and orientations, its ability to do so is hampered by the \textit{competition} from other garuds in the neighborhood that are also trying to do the same. As Venkatasubramanian explains \cite{venkat2017}, this \textit{disutility of competition} can be modeled as $-\delta \ln n_i$. 

This term, when integrated to obtain the potential $\phi(\mathbf{x})$, leads to entropy in statistical mechanics. Thus, maximizing the potential $\phi(\mathbf{x})$ is equivalent to maximizing entropy under certain conditions. For more details, the reader is referred to Venkatasubramanian \cite{venkat2017, venkatasubramanian2022unified}.  

Now, by combining all these components, we arrive at the \textit{effective utility} for the $i^{\text{th}}$ garud given by 

\begin{eqnarray}
    h_i &= \alpha n_i - \beta n_i^2+ {\gamma}\sum_{j} n_{ij} \mathbf{s}_i\cdot\mathbf{s}_j - \delta\ln n_i \\
    &=\alpha n_i - \beta n_i^2 + \gamma n_i l_i - \delta \ln n_i \label{eq:util}
\end{eqnarray}

where $l_i$ = $\frac{1}{n_i} \sum_{j} n_{ij}\mathbf{s}_i \cdot \mathbf{s}_j$ is the average alignment of agent $i$. Without any loss of generality, $\delta$ can be assumed to be 1, and will be done so for the rest of this paper. 

When $\alpha, \beta, \gamma=0$, the garuds don't have any preferences and hence fly randomly. This is what is captured by the remaining $-\ln n_i$ term, which we call  \textit{entropic restlessness}. 

Statistical teleodynamics, via potential game theory, proves that the self-organizing dynamics of the garuds would eventually result in an \textit{arbitrage equilibrium} where the effective utilities of all the garuds are the \textit{same}, i.e., $h_i = h^*$. 

In the next section, we discuss our simulation results that confirm this prediction.

\section{Results}

\begin{figure}[ht]
    \centering
    \includegraphics[width = \linewidth]{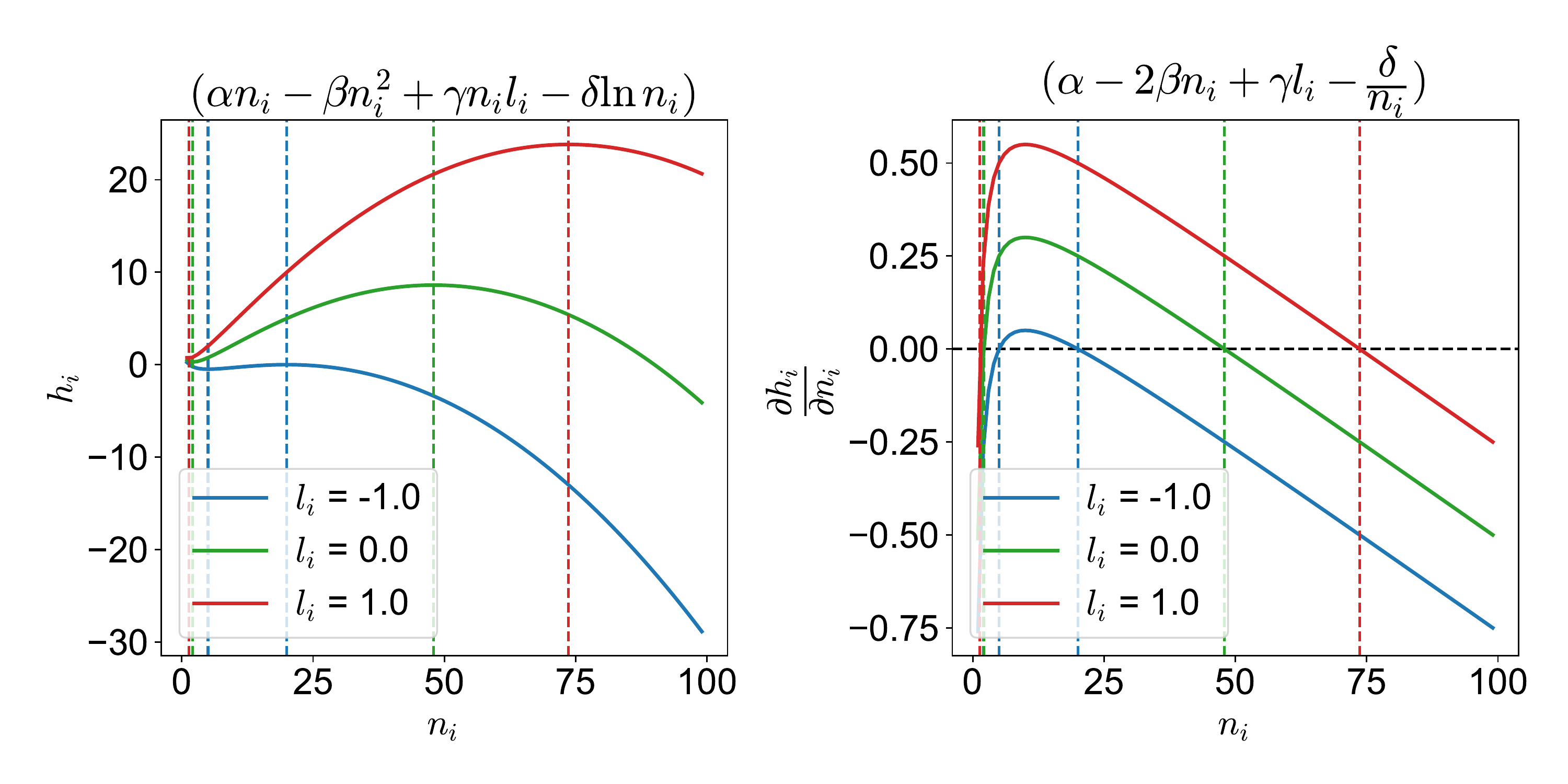}
    \caption{Effective utility and its derivative as a function of the number of neighbors $n_i$, for different values of alignment $l_i$ ($\alpha, \beta, \gamma, \delta = 0.5, 0.005, 0.25, 1$). There are two locations where the derivative of the effective utility is zero for different alignments.}
    \label{fig:utility}
\end{figure}

For the simulation details, the reader is referred to the methodology section VII below. If the garuds are flying randomly, without any rules of behavior, then this base case corresponds to $\alpha, \beta, \gamma, =0; \delta =1$. This result is discussed in Section S4. 

For the other cases, the effective utility function in Eq. \ref{eq:util} is plotted in Fig. \ref{fig:utility} in terms of the number of neighbors of garud $i$ ($n_i$), for different alignments ($l_i$)  for a given set of $\alpha, \beta, \gamma$. We see that there are two values of $n_i$ ($n_{-}$ and $n_{+}$) where the gradient of utility, for a given value of alignment, is zero. These values are determined by $\left(\dfrac{(\alpha+\gamma l_i) \mp \sqrt{(\alpha + \gamma l_i)^2 -8\beta\delta }}{4\beta}\right)$ (see, Supplementary Information, Section 1). 

In Fig. \ref{fig:utility}, at the lower value ($n_{-}$), any deviation in $n_i$ increases the utility of the garud, and hence leads to an \textit{unstable} point. However, for the higher $n_i$ point ($n_{+}$), we see that any deviation reduces the garud's utility. Therefore, this leads to a \textit{stable} point as any deviation would bring a garud back to the higher utility state. Therefore, this is the point a garud will try to reach to maximize its utility. For example, for the red curve, this would correspond to the point where $n_i = 73.6$. 

However, despite this point's stability, a garud will not be able to stay there indefinitely as the other garuds in its neighborhood are constantly changing their positions and orientations in their flights. Therefore, the $i^{\text{th}}$ garud would be fluctuating around this point. 

\begin{figure}[ht]
    \centering
    \includegraphics[width=\linewidth]{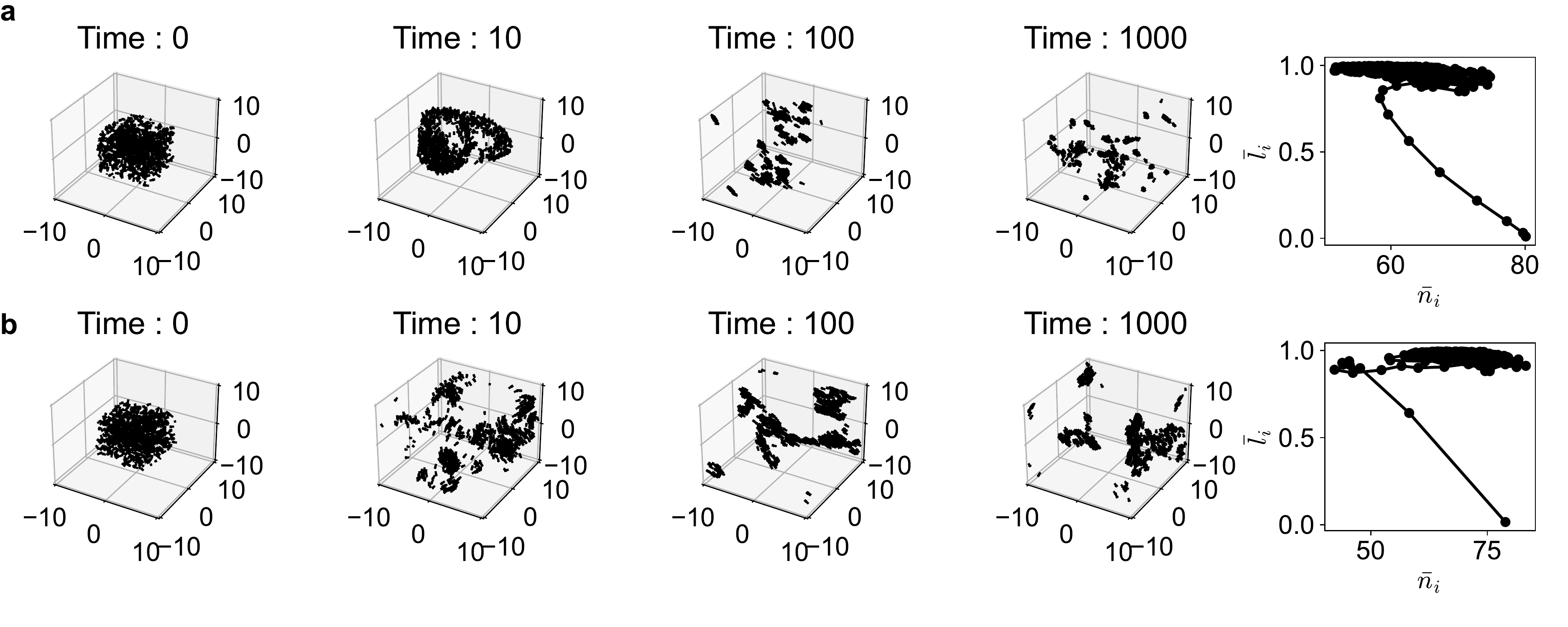}
    \caption{Trajectory of the agents and the corresponding phase portrait in $n_i$-$l_i$ space for the average number of neighbors of each individual agent during the course of the simulation, corresponding to (\textbf{a}) Reynolds's boids for $a=0.5, b=0.01, c=0.5$ and (\textbf{b}) Utility-driven garuds for $\alpha=0.5, \beta =0.01, \gamma=0.25, \delta=1$}
    \label{fig:pp}
\end{figure}

\begin{figure}[ht]
    \centering
    \includegraphics[width=\linewidth]{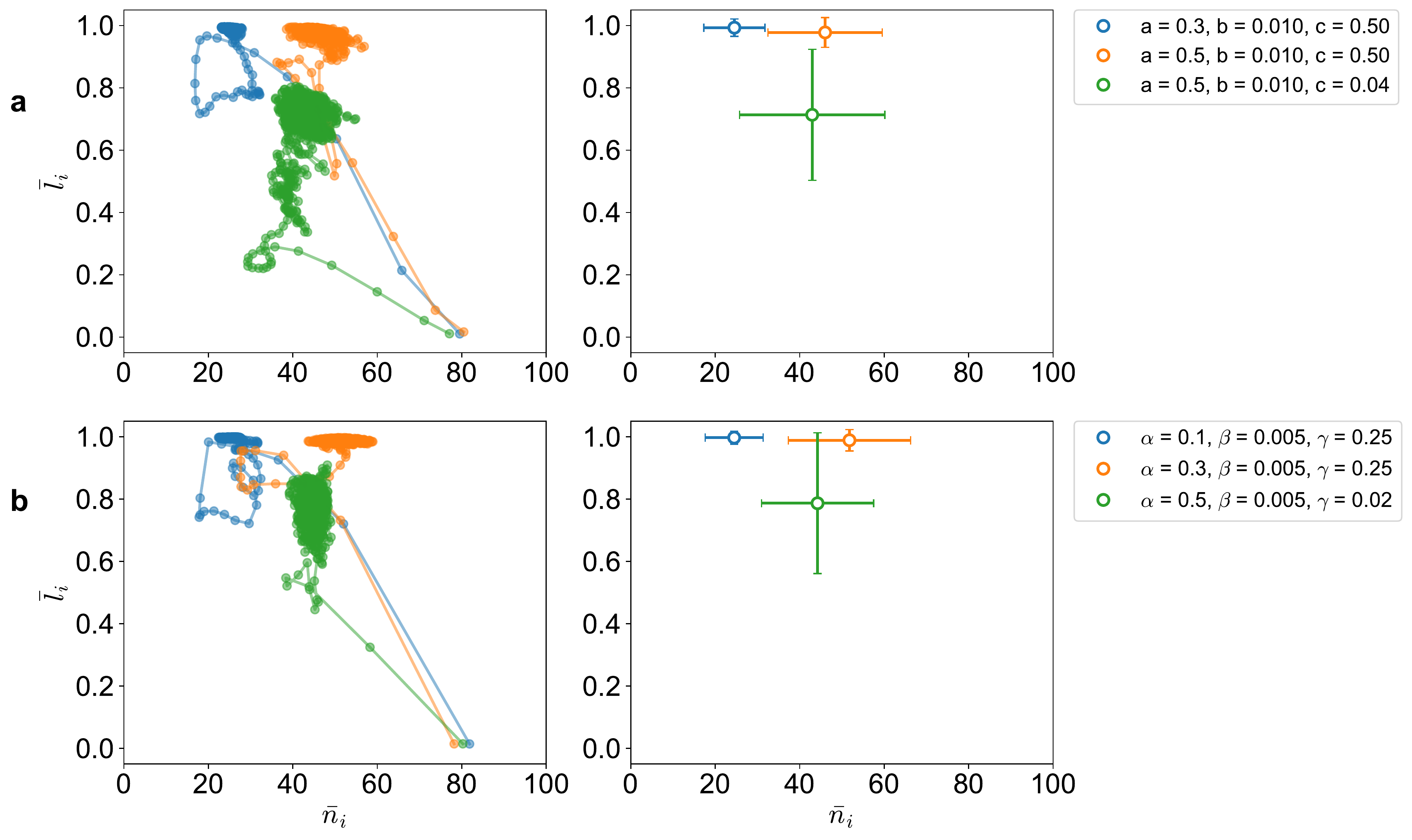}
    \caption{Trajectory of the average of number of neighbors of each agent and average alignment of the agents in the $n_i$, $l_i$ phase-space, and corresponding estimated averages for the (\textbf{a}) Reynolds's model (\textbf{b}) Utility-driven model ($\delta = 1$)}
    \label{fig:pp-all}
\end{figure}

In Fig.  \ref{fig:pp} and \ref{fig:pp-all}, we show the simulation results of both the Reynolds's \textit{Boids} model and our utility-driven \textit{Garud} model dynamics. We show the results of the Reynolds's model for illustrative purposes only as our objective is not to mimic the Reynolds's model exactly. We just want to show that the utility-driven model's collective behavior is very similar to that of the Reynolds model. 

The results are shown for different parameter values of ($a, b, c$) and ($\alpha, \beta, \gamma$). From the simulations, we obtain a set of position and velocity values $\{\mathbf{r}_i, \mathbf{v}_i\}$ of each agent $i$ at every time step. Once this is obtained, we extract the features $n_i$ and $l_i$ for all the agents. This in turn, is used to compute the average number of neighbors $\bar{n}$ and average alignment $\bar{l}$ for the entire population for all time points (Fig. \ref{fig:pp}). 

Figure \ref{fig:pp} shows the snapshots at different time points of the evolution in 3D-space and the $n_i$-$l_i$ phase-space. 

While the exact dynamics, the exact configuration of the population, and the time-scale of evolution of these two models cannot be the same (and that is not the aim, either), we observe that the qualitative patterns of collective behavior are  very similar in both cases. In particular, we see that all agents gravitate towards a certain region in the $n_i-l_i$ phase space for both models. They both start at the lower far right point (where the average alignment is near zero as the agents are all randomly oriented initially, and closely packed) and evolve towards the upper center-right region in black. We notice a qualitative match of both the trajectories. 


Furthermore, we can also see a quantitative similarity between the two models for specific parameters (Fig. \ref{fig:pp-all}). Fig. \ref{fig:pp-all}\textbf{a} shows the phase-space for the Reynolds's boids model, and Fig. \ref{fig:pp-all}\textbf{b} shows the same for the utility-driven model. 

In both models, we notice similar features of evolution towards the arbitrage equilibrium states, starting from the lower right point at time $t = 0$ to ending in the colored regions, where the average number of neighbors and the average alignment fall in similar corresponding regions. The plots in the right show the average alignment and average number of neighbors of the agents in the last 100 time steps. 

While both models exhibit similar collective behaviors, it is not apparent, however, from the three rules of the Reynolds's model, that its dynamics would result in an equilibrium state in the $n_i-l_i$ phase space. 

This is different for the utility-driven model, however. Since its potential game formulation predicts an arbitrage equilibrium outcome, it is clear right from the beginning where in the phase space the system is going to end up in. We can make a quantitative prediction about the average $n_i^*$ and $h^*$ values at equilibrium. 

This ability to predict the final outcome of the collective behavior of the population, given the individual agent-level properties captured in the utility function $h_i$, is an important defining feature and strength of the statistical teleodynamics framework. An additional characteristic is the ability to prove the stability of the final outcome.

\begin{figure}[ht]
    \centering
    \includegraphics[width=\linewidth]{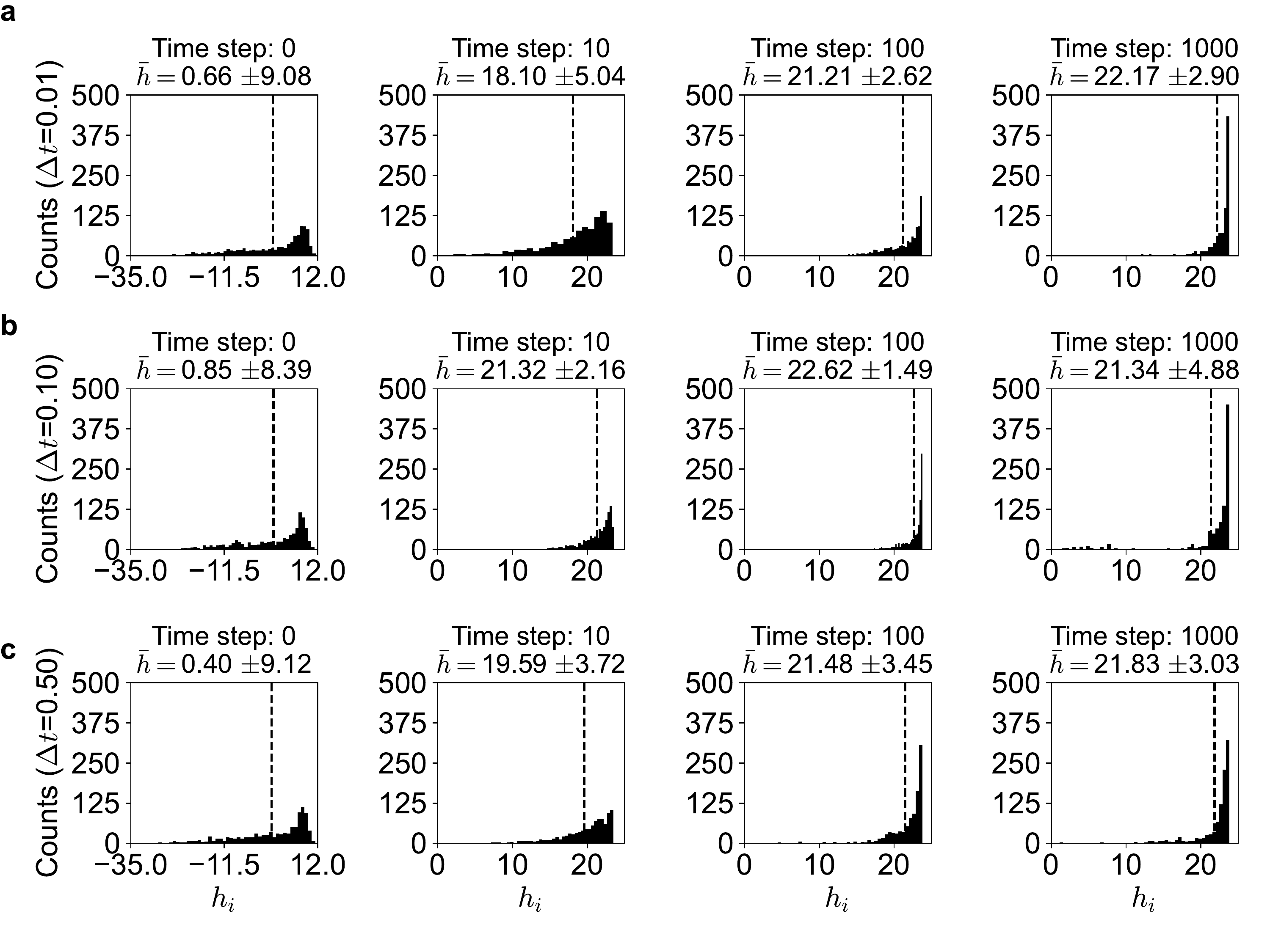}
    \caption{Histogram of the utility ($h_i$) of the garuds for $\alpha, \beta, \gamma, \delta = 0.5, 0.005, 0.25, 1$ corresponding to (\textbf{a}) $\Delta t = 0.01$, (\textbf{b}) $\Delta t = 0.1$ and (\textbf{c}) $\Delta t = 0.5$. Dashed line shows the average utility at a particular time.}
    \label{fig:utility-hist}
\end{figure}

We also ran the simulations for different time-step sizes of $\Delta t = 0.01, 0.1, 0.5$ in Eq. \ref{eq:boids_dt} to understand the dynamics of the evolution better. Note in Figure \ref{fig:utility-hist}, at the start ($t=0$), the utilities of all the garuds are spread out, with many having negative utility values, and the average utility ($\bar h$) low. 

But as the dynamics evolves, every garud tries to increase its utility by maneuvering to a better neighborhood and better orientation, the distribution becomes narrower, the average utility keeps increasing, and reaches a near-maximum value ($\bar h = 22.17 \pm 2.90$ in Fig. \ref{fig:utility-hist}a) and fluctuates around it. Note that this is around the maximum theoretical value of about 23.8 (given by Eq. \ref{eq:util}), where the histogram peaks. This suggests that nearly all the garuds have similar effective utilities asymptotically, approaching the maximum. This, of course, is the \textit{arbitrage equilibrium} outcome predicted by the theory (see also, Supplementary information, Fig. S3). 

The garuds do not converge exactly on $h^*$ but fluctuate around it because of the stochastic dynamics. This is also seen in Table \ref{tab:num-utility} where nearly the top 10 $\%$ of the garuds at a particular time-step are very close to the maximum utility value. In fact, the top $50 \%$ of the garuds have an average utility of greater than 23. 

This arbitrage equilibrium state is \textit{unique} only if the potential $\phi(\mathbf{x})$ is strictly \textit{concave} \cite{sandholm2010population}. For garuds, this is not the case in general as the concavity would depend on $\alpha$, $\beta$, and $\gamma$ having some particular values. So, for the typical case where $\phi(\mathbf{x})$ is not concave, there could be multiple equilibrium configurations of the garuds. Thus, instead of an equilibrium point in the $n_i-l_i$ phase space, one has an equilibrium region, in general. In other words, invoking a terminology from chaos and nonlinear dynamics, there would be a \textit{basin of attraction} in the phase space where the garuds finally settle in and fly around. This is what we see in Fig.  \ref{fig:pp} and \ref{fig:pp-all} in the colored regions. 

\begin{table}[]
    \centering
    \caption{Utility of different percentiles of the garuds at the $1000^\text{th}$ time step corresponding to Figure 4}
    \label{tab:num-utility}
    \begin{tabular}{|c|c|c|}
    \hline
        \textbf{Population}& \textbf{Time step size}, $\Delta t$ & \textbf{Average utility}\\ \hline
        \hline
         Top $1\%$ & 0.01 & 23.79\\
         & 0.1 & 23.80\\
         & 0.5 & 23.76\\ \hline
         Top $1-10\%$ & 0.01 & 23.77\\
         & 0.1 & 23.78\\
         & 0.5 & 23.68\\ \hline
         Top $10-50\%$ & 0.01 & 23.60\\
         & 0.1 & 23.51\\
         & 0.5 & 23.35\\ \hline
         Top $50-75\%$ & 0.01 & 22.83\\
         & 0.1 & 22.64\\
         & 0.5 & 22.58\\ \hline
          Bottom $50\%$ & 0.01 & 18.59\\
         & 0.1 & 15.59\\
         & 0.5 & 17.91\\ \hline
    \end{tabular}
\end{table}



\subsection{Stability of the Arbitrage Equilibrium}

We can ascertain the stability of this equilibrium by performing a Lyapunov stability analysis \cite{venkat2017}. A Lyapunov function $V$ is a continuously differentiable function that
takes positive values everywhere except at the equilibrium point (i.e., $V$ is positive definite), and decreases (or is nonincreasing) along every
trajectory traversed by the dynamical system ($\dot{V}$ is negative definite or negative semidefinite). A dynamical system is \textit{locally stable} at equilibrium if $\dot{V}$ is negative semidefinite and is \textit{asymptotically stable} if $\dot{V}$ is negative definite.

Following Venkatasubramanian~\cite{venkat2017}, we identify our Lyapunov function $V(n_i)$

\begin{eqnarray}
V(n_i) = \phi^*(n_i^*) - \phi(n_i) \label{eq:lyapunov}
\end{eqnarray}

where $\phi^*$ is the potential at the Nash equilibrium (recall that $\phi^*$ is at its maximum at NE) and $\phi(n_i)$ is the potential at any other state. Note that $V(n_i)$ has the desirable properties we seek: (i) $V(n^*)$ = 0 at NE and $V(n_i)$ $>$ 0 elsewhere, i.e., $V(n_i)$ is positive definite; (ii) since $\phi(n_i)$ increases as it approaches the maximum, $V(n_i)$ decreases with time, and hence it is easy to see that $\dot{V}$ is negative definite. Therefore, the arbitrage equilibrium is not only \textit{stable} but also \textit{asymptotically stable}. 

\begin{figure}[ht]
    \centering
    \includegraphics[width=\linewidth]{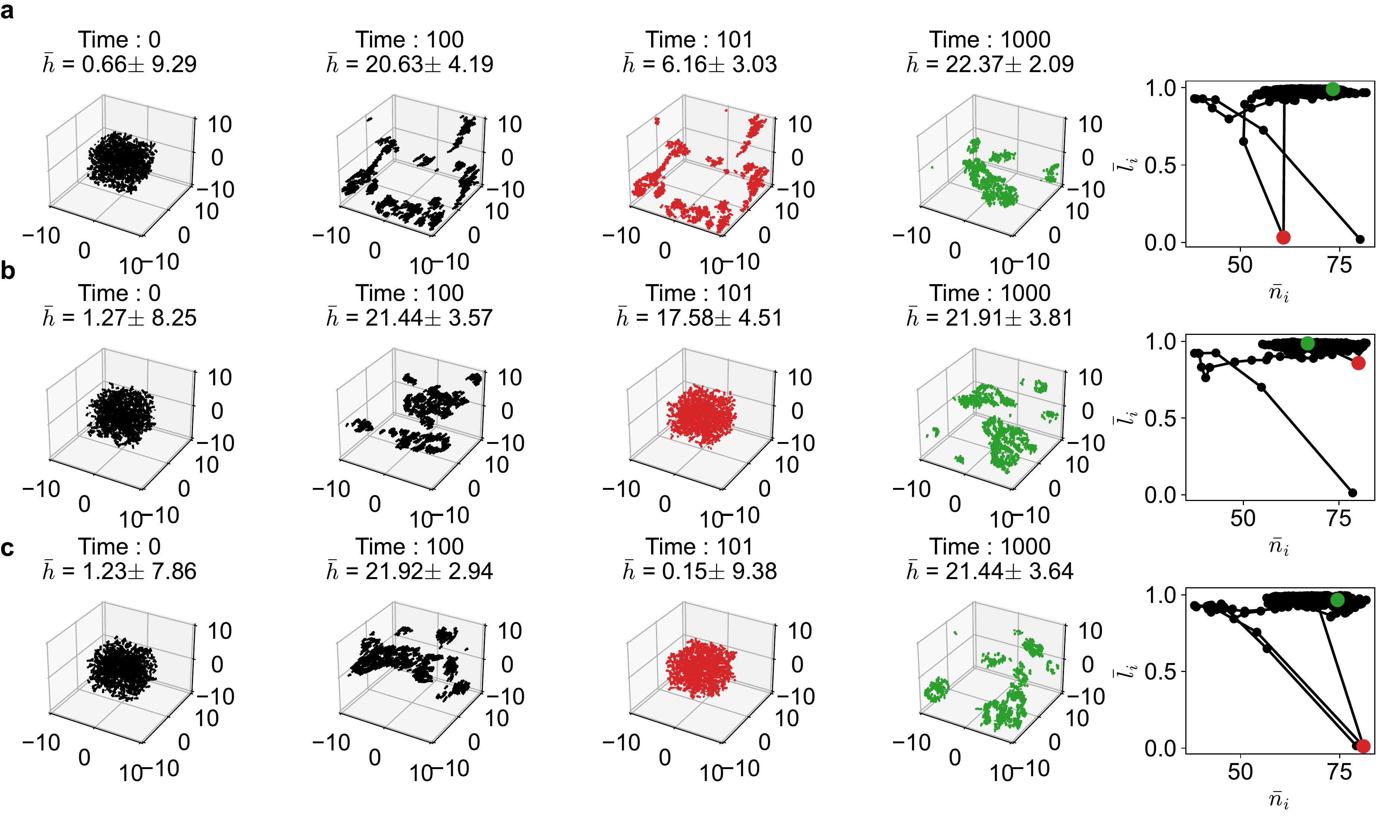}
    \caption{The trajectory of the garuds as a function of time and the corresponding phase-portraits for $\alpha, \beta, \gamma, \delta = 0.5, 0.005, 0.25, 1$ for stability analysis case studies (\textbf{a}) Disturbance 1, (\textbf{b}) Disturbance 2, and (\textbf{c}) Disturbance 3. The disturbances occur after equilibrium at time step 101.}
    \label{fig:deviation}
\end{figure}

Our simulation results confirm this theoretical prediction (see Figure \ref {fig:deviation}). After the garuds population reached equilibrium, we disturbed the equilibrium state by randomly changing the positions and/or velocities of the garuds. The simulation is then continued from the new disturbed far-from-equilibrium state. We conducted experiments with three kinds of disturbances:

\begin{itemize}
    \item \textbf{Disturbance 1}:\textit{Velocity disturbance}, where each garud's velocity is changed to a random orientation and magnitude.
    \item \textbf{Disturbance 2}: \textit{Position disturbance}, where each garud's position is randomly changed.
    \item \textbf{Disturbance 3}: \textit{Position and velocity disturbance}, where both position and velocity vectors are changed.
\end{itemize}

As seen in Figure \ref{fig:deviation}, after the 100$^\text{th}$ time step when the population had reached equilibrium, we introduced these disturbances. The 101$^\text{st}$ time step shows in red color the new far-from equilibrium states, where the average utility has dropped considerably. 

In all cases, the population recovers quickly, typically in another 100 time steps or so to reach the original equilibrium region (shown in green). The figure shows the disturbance (red) and recovery (green) in both the 3-D space and the phase space. 

In Fig. \ref{fig:deviation}\textbf{a}, as the velocities are randomized at the 101$^\text{st}$ time step, the alignment goes down to 0, but recovers to the original equilibrium quickly. In Fig. \ref{fig:deviation}\textbf{b}, as the new configuration corresponds to a similar value of average number of neighbors as before, the disturbance is not that much. Note that the drop in average utility is small. In Fig. \ref{fig:deviation}\textbf{c}, we see that this disturbance is huge, pushing the configuration close to the original random state. But still, the population is able to recover to the arbitrage equilibrium quickly. 

This shows that the arbitrage equilibrium region is not only stable, but asymptotically stable. {That is, the garuds flocking configuration is \textit{resilient} and \textit{self-healing}.} Given the speed of the recovery, it could possibly be exponentially stable but we have not proved this analytically here. 

The asymptotic stability of this arbitrage equilibrium is similar to that of the income-game dynamics as discussed by Venkatasubramanian \cite{venkat2015, venkat2017} using a similar Lyapunov stability analysis.

\section{Conclusion}

For three centuries we have known that there are constants of motion, such as energy and momentum, for passive matter.  Nevertheless, it comes as a surprise to discover that the dynamics of active matter populations could also have an invariant, namely, the effective utility. However, the role of invariance here is different from its role in dynamics. The constants of motion such as energy and momentum are conserved, but effective utility is not. 

The role of this invariance is more like that of \textit{set point} tracking and disturbance rejection in feedback control systems. These are called the regulation and servo problems, respectively, in control theory \cite{seborg2016process, aastrom2021feedback}. The system, i.e., the garuds population, adjusts itself dynamically and continually, in a feedback control-like manner, to maintain its overall effective utility. 

It is important to emphasize, however, that this control action is decentralized as opposed to the typical centralized control system in many engineering applications. The agents individually self-organize, adapt, and dynamically course-correct to offset the negative impact on their effective utilities by other agents or other external sources of disturbance. The population as a whole stochastically evolves towards the stable \textit{basin of attraction} in the phase space in a self-organized and distributed-control fashion. 

{This is essentially Adam Smith's \textit{invisible hand} mechanism of economics. As Smith observed\cite{smith1776wealth}, "It is not from the benevolence of the butcher, the brewer, or the baker, that we expect our dinner, but from their regard to their own interest. We address ourselves, not to their humanity but to their self-love, and never talk to them of our own necessities but of their advantages." Thus, every garud is pursuing its own self-interest, to increase its own $h_i$, and a stable collective order spontaneously emerges via such self-organization.}

Invariants are quite rare in physics, rarer still in biology and economics. That's why it is exciting to see them as their presence usually signals something deep, something fundamental. In physics, their existence has revealed deeply fundamental symmetries of the cosmos, as Emmy Noether showed. Therefore, it is important to understand the implications of our discovery in sufficient depth and breadth. 

We do realize, of course, that our bird-like agents are not real birds. Our model and simulations are not real biological systems. Nevertheless, our results suggest intriguing possibilities for real biological entities that need to be explored carefully. 

Therefore, the interesting and surprising result, seen both analytically and in simulations, that the emergent arbitrage equilibrium is asymptotically stable is an important one with potentially far-reaching consequences, particularly in biological, ecological, and economic contexts. For example, this could be an important mechanism of pattern formation and pattern stability in biological systems. Populations of cells could self-organize, under different spatial and temporal conditions and constraints, driven by their incessant and instinctive hunt for better utilities, to settle into various stable basins of attraction - i.e., into different types of stable emergent order - to form stable organized structures. {Their asymptotic stability property bestows upon them the \textit{resilient self-healing} feature found so commonly in many biological systems. This process could be a core mechanism behind the design, control, and optimization of stable biological systems via self-organization.} 

This mechanism is applicable to different length and time scales, from molecular to macroscopic to planetary scales. These results raise several interesting questions about populations of biological active matter competing with one another. For instance, consider all the different kinds of microbial populations in the human body, or for that matter, in any living organism. Not only are all the individual microbes competing with one other strategically for resources to increase their effective utilities, to improve their survival and growth fitness, different microbial populations are also competing with one another at a higher-scale. 

So, is there a hierarchy of arbitrage equilibria? That is, the microbes in a given population at some arbitrage equilibrium among themselves (say, level-1 equilibrium), and such populations themselves are in equilibrium with one another at a higher level (say, level-2), and so on. Is there a planet-scale equilibrium? That is, are all the living species on our planet, along with the environment, at some arbitrage equilibrium?  Or are we evolving towards one, just like the garuds population did in this study. What happens when this equilibrium is upset by either internal disturbances (such as climate change) or external shocks (such as asteroid impact)?

As one can see, our theory is not limited to just birds flocking. It is also applicable to the self-organizing dynamics and evolution of a wide variety of systems in physics, biology, sociology, and economics. As Venkatasubramanian et al.\cite{venkatasubramanian2022unified} showed the emergence of the exponential energy (i.e., Boltzmann) distribution for gas molecules can be modeled by the effective utility 

\begin{equation}
h_{i}= -\beta \ln E_i - \ln N_i.
\end{equation}

Similarly, they showed,\cite{venkatasubramanian2022unified} as examples of biological systems, bacterial chemotaxis can be modeled by 

\begin{equation}
h_{i}= -\alpha c_i - \ln N_i
\end{equation}

and the emergence of ant craters by 

\begin{equation}
    h_i = b - \frac{\omega r_i^a}{a} - \ln N_i. 
\end{equation}

The same study showed how the Schelling game-like segregation dynamics in sociology can be modeled by
\begin{eqnarray}
    h_i = \eta N_i+ \ln(H - N_i) - \xi N_i^2 - \ln N_i
\end{eqnarray}

and the income game in economics by 

\begin{equation}
h_{i}= \alpha \ln S_i - \beta \left(\ln S_i\right)^2 - \ln N_i.
\end{equation}

What we have is for ideal systems, similar to the ideal gas or Ising model in thermodynamics. Just as real gases and liquids don't behave exactly like their ideal versions in statistical thermodynamics, we don't expect real biological systems (or economic or ecological systems) to behave like their ideal counterparts in statistical teleodynamics. Nevertheless, the ideal versions serve as useful starting and reference points as we develop more comprehensive models of active matter systems.  The next steps would involve examining and learning how real-world biological systems behave compared to their ideal versions. This would, of course, necessitate several modifications to the ideal models. 

We note, from equations 14-18, a certain pattern in the structure of the effective utility functions in different domains. Thus, we see that the same mathematical and conceptual framework is able to predict and explain the emergence of {spontaneous} order via self-organization to reach arbitrage equilibrium in dynamical systems in physics, biology, sociology, and economics. 

This kind of universality is particularly striking, prompting us to conclude with a quote a from the inimitable Richard Feynman that seems \textit{apropos} here: "Nature uses only the longest threads to weave her patterns, so that each small piece of her fabric reveals the organization of the entire tapestry." It appears that the emergence of spontaneous order via self-organizing stable arbitrage equilibria is such a thread.

\section{Acknowledgements}
This work was supported in part by the Center for the Management of Systemic Risk (CMSR), Columbia University. This manuscript was written when the corresponding author (VV) was the Otto Monsted Distinguished Visiting Professor at the Danish Technical University (DTU) as well as a resident of Nyhavn 18, Copenhagen, as a guest of the Danish Central Bank in the summer of 2022. It is with great pleasure that VV acknowledges the generous support of these institutions and the warm hospitality of his colleagues in the Chemical Engineering Department at DTU. 

\section{Authorship contribution statement} 
Venkat Venkatasubramanian: Conceptualization, Formal analysis, Funding acquisition, Methodology, Investigation, Supervision, Writing – original draft, Writing –review \& editing. 

Abhishek Sivaram: Software, Formal analysis, Writing –review \& editing, Data curation, Validation. 

\section{Methodology}
We created a simulation of 1000 garuds in a periodic-box of dimensions $20\times 20 \times 20$, where each garud's neighborhood is a sphere with radius $r_0 = 3$. Each garud starts at a random location and orientation inside a $10\times 10 \times 10$ block. Speed of each garud is limited between $0.5$ and $1$. The update algorithm works similar to the Reynolds's garuds update, except the force is driven by the numerical estimates of direction of increased utility (additively based on position and velocity). An additional noise is also added to the velocity update strategy similar to the Reynolds's model to capture the erroneous strategies of velocity update for each garud. This is given by a noise parameter (0.01, unless specified) times the magnitude of the velocity. The noise indicates that a garud does not make perfect choices in updating its velocity.

\bibliographystyle{apsrev4-2}

%

\end{document}


\title{\textsc{Supplementary Information}\\Arbitrage Equilibrium, Invariance, and the Emergence of Spontaneous Order\\ in the Dynamics of Birds Flocking}

\author{Abhishek Sivaram}
\affiliation{Complex Resilient Intelligent Systems Laboratory, Department of Chemical Engineering, Columbia University, New York, NY 10027, U.S.A.}
\author{Venkat Venkatasubramanian}%
 \email{venkat@columbia.edu}
\affiliation{%
Complex Resilient Intelligent Systems Laboratory, Department of Chemical Engineering, Columbia University, New York, NY 10027, U.S.A.
}%
\maketitle

\maketitle

\section{Utility Model for Birds Flocking}
\label{sec:1}
We can see that the number of neighbors for a garud $n_i = \sum_j n_{ij}$. The garuds try to maximize the effective utility, given by 
\begin{eqnarray*}
    h_i &&= \alpha n_i - \beta n_i^2 + \gamma \sum_{j} n_{ij} \mathbf{s}_i \cdot \mathbf{s}_j - \delta \ln n_i
\end{eqnarray*}

Here $\mathbf{s}_i = \dfrac{\mathbf{v}_i}{|\mathbf{v}_i|}$. We define the average alignment of each garud as $l_i = \dfrac{\sum_j n_{ij} \mathbf{s}_i \cdot \mathbf{s}_j}{n_i}$.

This results in the alternative formulation of the utility in terms of the alignment, 
\begin{eqnarray*}
h_i = \alpha n_i - \beta n_i^2 +\gamma n_i l_i - \delta \ln n_i
\end{eqnarray*}

The first three terms are the utility of cohesion, disutility of congestion, and the utility of alignment. The last term is the disutility due to entropic restlessness.

\section{Optimum Based on the utility formulation}
\label{sec:2}
At equilibrium, all garuds  have the same utility, i.e.,

\begin{eqnarray*}
h_i = \alpha n_i - \beta n_i^2 +\gamma n_i l_i - \delta \ln n_i = h^*
\end{eqnarray*}

Now, $h_i$ is maximum at two $n_i$ values where the gradient is zero, given by, 
\begin{eqnarray*}
    \frac{\partial h_i}{\partial n_i}  = \alpha - 2\beta n_i + \gamma l_i - \frac{\delta}{n_i}= 0 \\
    \implies - 2\beta n_i^2 + (\alpha + \gamma l_i) n_i  - \delta &&=0\\
    n_{\mp} = \frac{(\alpha+\gamma l_i) \mp \sqrt{(\alpha + \gamma l_i)^2 -8\beta\delta }}{4\beta} 
\end{eqnarray*}

Note that $n_{-}$ is an \textit{unstable} point as any deviation in the number of neighbors would result in increasing utility, thereby causing the garud to move away from there. On the other hand, $n_{+}$ is a \textit{stable} point because any deviation would decrease the utility, thereby causing the garud to return to its original state. 

\section{Dynamical Models of Flocking}
\label{sec:3}

Regarding the discussion in Section II, the net effect of the three forces in the Reynolds model on the velocity of the $i$th boid is modeled by the equation, 

\begin{eqnarray}
    \mathbf{v}_i(t+1) = \mathbf{v}_i(t) && + a (\mathbf{r}_{c,i} - \mathbf{r}_i)  + b \sum_{j} n_{ij}(\mathbf{r}_i -\mathbf{r}_j) \nonumber\\ && + c (\mathbf{v}_{c,i} - \mathbf{v}_i)  + \boldsymbol{\eta}(t) \label{eq:boids}
\end{eqnarray}

where $a, b,$ and $c$ are parameters corresponding to the \textit{rule of cohesion}, \textit{rule of separation}, and \textit{rule of alignment}, respectively, $\mathbf{v}_{c,i}$ is the average velocity of the neighbors of $i$, and $\mathbf{r}_{c,i}$ is the center of the neighborhood as perceived by the agent $i$~\cite{boids-url}. These are given by

\begin{eqnarray*}
    \mathbf{v}_{c, i} &&= \frac{1}{n_i}\sum_{j\in \mathcal{N}^i} \mathbf{v}_j = \frac{1}{n_i}\sum_{j} n_{ij} \mathbf{v}_j\\
    \mathbf{r}_{c, i} &&= \frac{1}{n_i}\sum_{j} n_{ij}\mathbf{r}_j\\
\end{eqnarray*}
Parameter $\boldsymbol{\eta}$ is the uncorrelated noise in the agent's velocity. Substituting the average velocity of the neighbors and the center of the neighborhood as perceived by the agent $i$, Eq. \ref{eq:boids} can be simplified to give, 
\begin{eqnarray}
    \mathbf{v}_i(t+1) = \mathbf{v}_i(t) &&+\frac{a}{n_i} \sum_{j} n_{ij} (\mathbf{r}_j - \mathbf{r}_i) + b \sum_{j} n_{ij} (\mathbf{r}_i - \mathbf{r}_j) \nonumber\\&&  + \frac{c}{n_i} \sum_{j} n_{ij} (\mathbf{v}_j - \mathbf{v}_i) + \boldsymbol{\eta}(t) \label{eq:boids2}
\end{eqnarray}

In general, we can write the above equation as, 
\begin{widetext}
\begin{eqnarray}
    \mathbf{v}_i(t+\Delta t) = \mathbf{v}_i(t) + \left(\frac{a}{n_i} \sum_{j} n_{ij} (\mathbf{r}_j - \mathbf{r}_i) + b \sum_{j} n_{ij} (\mathbf{r}_i - \mathbf{r}_j)  + \frac{c}{n_i} \sum_{j} n_{ij} (\mathbf{v}_j - \mathbf{v}_i) + \boldsymbol{\eta}(t)\right) \Delta t \label{eq:boids_dt}
\end{eqnarray}
\end{widetext}

for a time-step $\Delta t$. The time-scale $\Delta t$ in Eq. \ref{eq:boids_dt} can be subsumed by the parameters to give Eq. \ref{eq:boids2}




The Vicsek model is a similar model where the velocity update is purely a function of the alignment of an agent with it's neighbors. The constant velocity dynamics is sometimes modified to include other pair-wise attraction-repulsion forces $\mathbf{f}_{ij}$ and is written as \cite{bialek2012statistical},

\begin{eqnarray*}
    \mathbf{r}_i(t+1) &&= \mathbf{r}_i(t) + \mathbf{v}_i(t)\\
    \mathbf{v}_i(t+1) &&= v_0\Theta\left[\alpha\sum_j n_{ij} \mathbf{v}_j(t) + \beta \sum_{j} n_{ij}\mathbf{f}_{ij} + \eta_i(t)\right]
\end{eqnarray*}

where $\Theta$ scales the dynamics to a unit vector to ensure the constant speed of the individual agents.

\section{Base case: $\alpha, \beta, \gamma = 0$}
\label{sec:4}
\begin{figure}
    \centering
    \includegraphics[width=\linewidth]{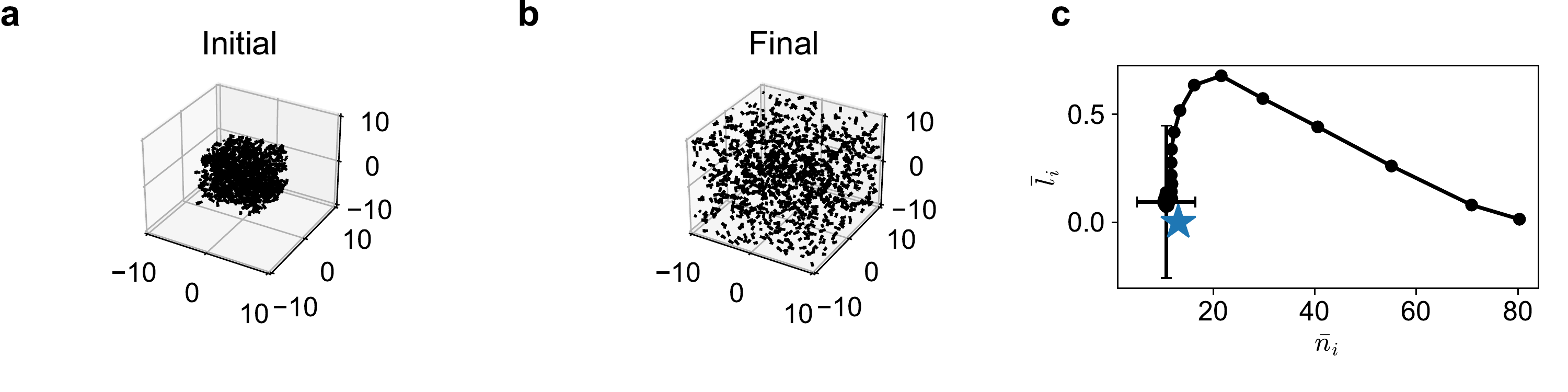}
    \caption{(\textbf{a}) Initial configuration of garuds, (\textbf{b}) Final configuration, and (\textbf{c}) Phase space for $\alpha, \beta, \gamma = 0, 0, 0; \delta =1$. The system is entropically driven. The star shows the theoretical average $\bar n_i$, $\bar l_i$ in the case where all the garuds are randomly distributed ($\bar n_i$ = $\sim$13, $\bar l_i$=0).}
    \label{fig:base}
\end{figure}

Fig. \ref{fig:base} shows the case where $\alpha, \beta, \gamma$ are set to zero, so the garuds are entropically driven (only $-\ln n_i$ component is driving the motion of the garuds). In this case, the garuds start with a high density and zero alignment (Fig. \ref{fig:base}\textbf{a}) initial configuration and finally settle in a configuration where they have random positions and velocities (Fig. \ref{fig:base}\textbf{b}). If each garud is randomly located in 3D space, the number of neighbors for each garud on average is given by,
\begin{eqnarray}
 \bar{n}_i = \frac{N}{L^3} \left(\frac{4}{3} \pi r_0^3\right) - 1 \label{eq:rand}
\end{eqnarray}
where $L = 20$ is the length of the domain and $r_0 = 3$  is the size of the neighborhood. We subtract 1 in Eq. \ref{eq:rand}, as the $i^\text{th}$ garud itself is not considered in the average number of neighbors. With a total of 1000 boids, this gives an estimate of the average number of neighbors of the $i^\text{th}$ boid as 13.1. The random alignment of the garud also gives the average alignment with neighbors as zero, as there is no incentive to increasing the alignment. The simulation results confirm the theoretical expectations. 

\section{Noise trend}
\label{sec:5}
\begin{figure}
    \centering
    \includegraphics[width=\linewidth]{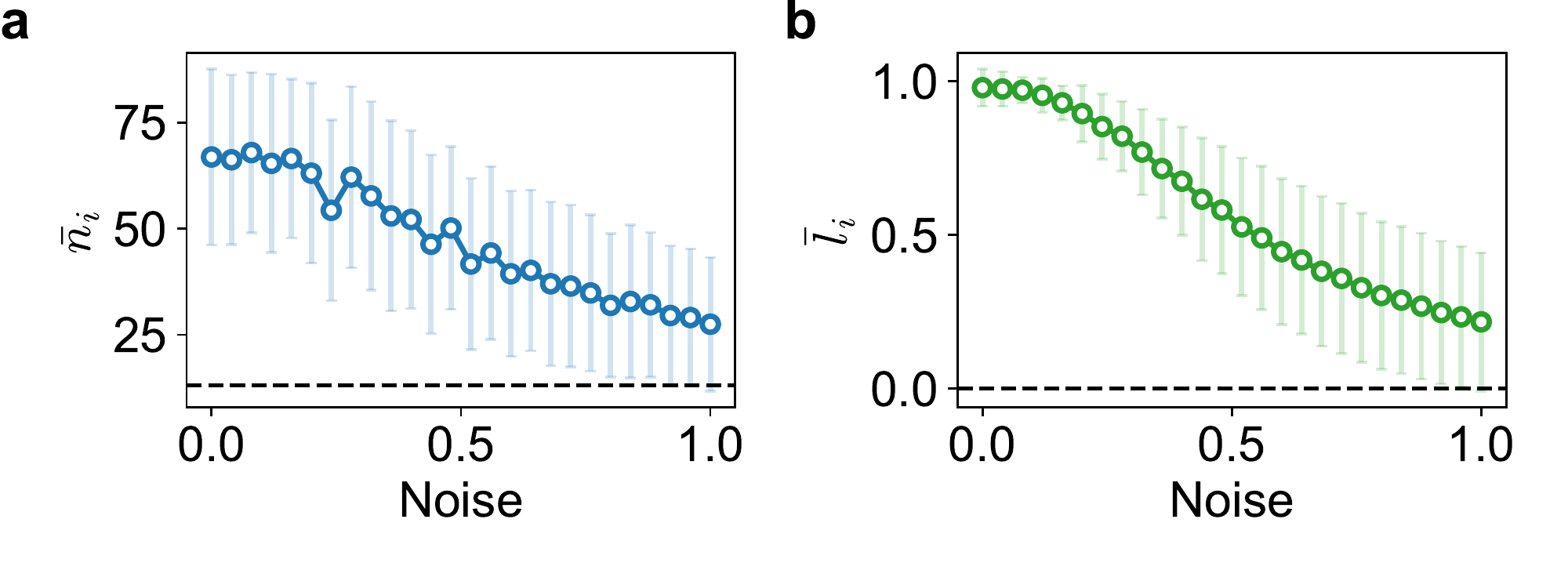}
    \caption{Neighbors (\textbf{a}) and alignment (\textbf{b}) dependence on noise parameter for utility parameters $\alpha, \beta, \gamma = 0.5, 0.005, 0.25$. We see that increasing the noise in decision-making results in the system tending more towards random behavior (dashed lines in \textbf{a} and \textbf{b}) }
    \label{fig:noise}
\end{figure}

The noise parameter is also varied in the simulations. To reiterate, the noise-parameter dictates the magnitude of randomness in the velocity vector which is added to the ideal ``direction'' of increasing utility. As the noise keeps increasing, we see that the system tends more and more towards randomness characterized by random positions and velocities of agents (dashed lines in Fig. \ref{fig:noise}). 

\section{Dynamics variation for different time step sizes}
\label{sec:6}
\begin{figure}
    \centering
    \includegraphics[width=\linewidth]{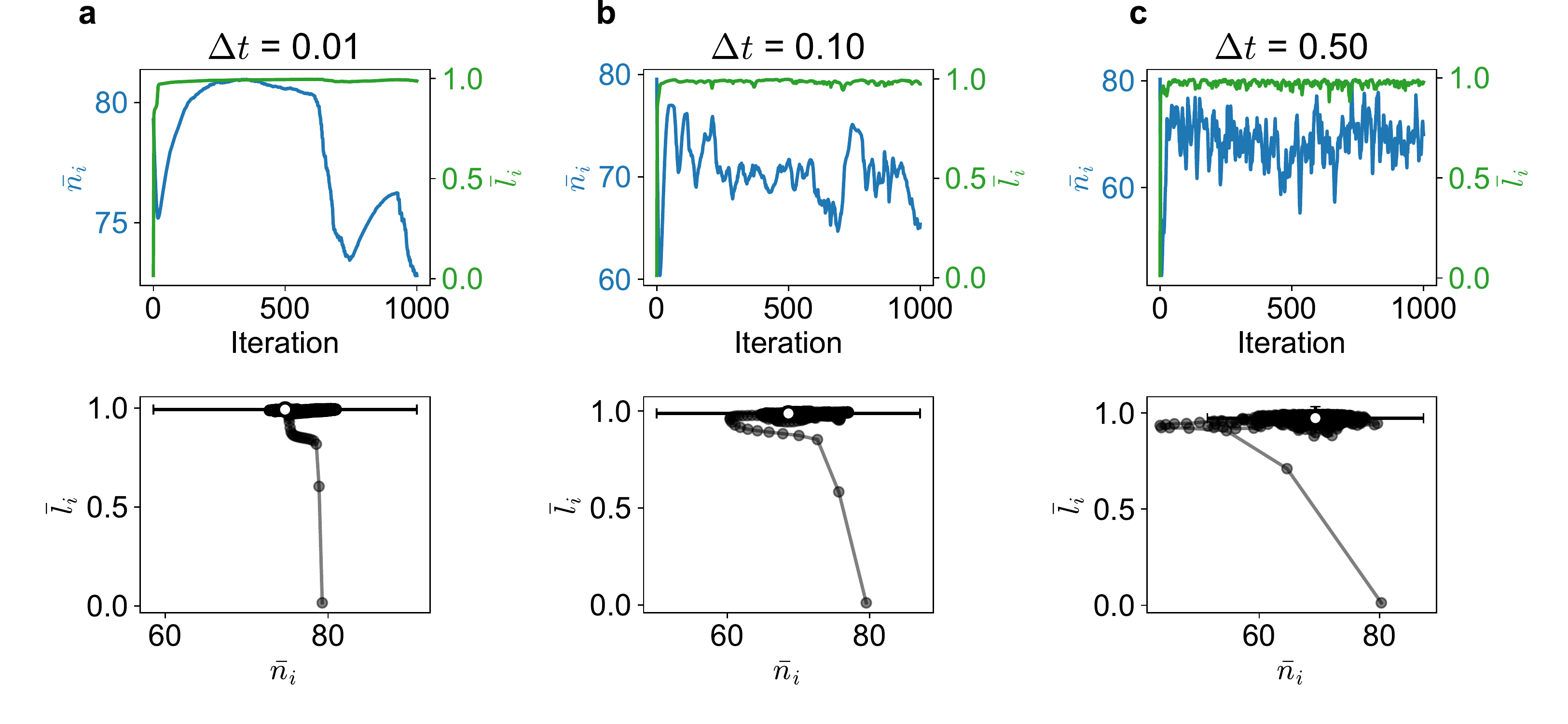}
    \caption{The trajectory of average number of neighbors $\bar{n}$ and average alignment $\bar{l}$ as function of iteration time for step-sizes (\textbf{a}) $\Delta t = 0.01$, (\textbf{b}) $\Delta t = 0.1$, and (\textbf{c}) $\Delta t = 0.5$ along with the corresponding phase-space plots (below) for $\alpha, \beta, \gamma, \delta = 0.5, 0.005, 0.25, 1$. .}
    \label{fig:delta-t}
\end{figure}

We also ran the simulation for different time step sizes of $\Delta t = 0.01, 0.1, 0.5$ in Eq. \ref{eq:boids_dt}. Fig. \ref{fig:delta-t} shows that the dynamics are quite different of the garuds for the different step sizes. However, the corresponding phase plots suggest that irrespective of the different dynamics, the garuds return to the same region in the phase space at equilibrium. 


\bibliographystyle{apsrev4-2}